# An Empirical Study of C++ Vulnerabilities in Crowd-Sourced Code Examples

Morteza Verdi, Ashkan Sami, Jafar Akhondali, Foutse Khomh, Gias Uddin, and Alireza Karami Motlagh

**Abstract**—Software developers share programming solutions in Q&A sites like Stack Overflow, Stack Exchange, Android forum, and so on. The reuse of crowd-sourced code snippets can facilitate rapid prototyping. However, recent research shows that the shared code snippets may be of low quality and can even contain vulnerabilities. This paper aims to understand the nature and the prevalence of security vulnerabilities in crowd-sourced code examples. To achieve this goal, we investigate security vulnerabilities in the C++ code snippets shared on Stack Overflow over a period of 10 years. In collaborative sessions involving multiple human coders, we manually assessed each code snippet for security vulnerabilities following CWE (Common Weakness Enumeration) guidelines. From the 72,483 reviewed code snippets used in at least one project hosted on GitHub, we found a total of 99 vulnerable code snippets categorized into 31 types. Many of the investigated code snippets are still not corrected on Stack Overflow. The 99 vulnerable code snippets found in Stack Overflow were reused in a total of 2859 GitHub projects. To help improve the quality of code snippets shared on Stack Overflow, we developed a browser extension that allows Stack Overflow users to be notified for vulnerabilities in code snippets when they see them on the platform.

✦

## 1 INTRODUCTION

A major goal of software development is to deliver high quality software in timely and cost-efficient manner. Code reuse is an accepted practice and an essential approach to achieve this premise [1]. The reused code snippets come from many different sources and in different forms, e.g., third-party library [2], open source software [3], and Question and Answer (Q&A) websites such as Stack Overflow [4], [5]. Sharing code snippets and code examples is also a common learning practice [6]. Novices and even more senior developers leverage code examples and explanations shared on platforms like Stack Overflow, to learn how to perform new programming tasks or use certain APIs [1], [7], [8], [9]. Multiple studies [10], [11], [12] have investigated knowledge flow and knowledge sharing from Stack Overflow answers to repositories of open source software hosted in GitHub. They report that code snippets found on Stack Overflow can be toxic, i.e., of poor quality, and can potentially lead to license violations [12], [13]. An important aspect of quality that has not been investigated in details by the research com-munity is security. If vulnerable codes snippets are migrated from Stack Overflow to applications, these applications will be prone to attacks.

M. Verdi is with Shiraz University, Iran. E-mail: m.verdi@shirazu.ac.ir
A. Sami (Corresponding Author) is with Shiraz University, Iran. E-mail: sami@shirazu.ac.ir
J. Akhondali is with Shiraz University, Iran. E-mail: jafar.akhondali@yahoo.com
F. Khomh is with Polytechnique Montreal, Quebec Canada. E-mail: foutse.khomh@polymtl.ca
G. Uddin is with University of Calgary, Alberta Canada. Email: gias.uddin@ucalgary.ca
A. Karami Motlagh is with Shiraz University, Iran. E-mail: alireza.karami.m@gmail.com

The danger of copy pasting insecure code from Stack Overflow was recently raised by Fischer et al. [8], who found that vulnerable Android code snippets from Stack Overflow are reused in popular Android apps. We are, however, aware of no previous study that specifically fo-cused on the vulnerability of C++ code snippets shared in Stack Overflow and whether and how such vulnerable code snippets may have migrated to open source software repositories in GitHub. This insight is important because such vulnerable software repositories then can be reused by other software repositories, which given the popularity of GitHub, is entirely possible. C++ is the fourth most popular programming language [14]. C++ is the language of choice for embedded, resource-constrained programs. It is exten-sively used in large and distributed systems. Vulnerabilities in C++ code snippets are therefore likely to have a major impact. However, to the best of our knowledge, no study has examined the security aspects of C++ Stack Overflow code snippets and their impact on open source software projects. This paper aims to fill this gap in the literature. More specif-ically, we aim to understand the nature and the prevalence of security vulnerabilities in code examples shared on Stack Overflow. To achieve this goal, we empirically study C++ vulnerabilities in code examples shared on Stack Overflow along the following two dimensions.

Prevalence:] We review the C++ vulnerability types con-tained in a Stack Overflow data-set named SOTorrent [15], [16] and analyze their evolution over time; in particular their migration to GitHub projects.
From 72,483 C++ code snippets reused in at least one GitHub project, we found 99 vulnerable code snippets belonging to 31 different types of vulnerabilities.

**Propagation:** We investigate how the vulnerable code snip-pets were reused in GitHub repositories.

TABLE 1: Research contributions made in this paper to understand the prevalence and propagation of C++ vulnerabilities in crowd-Sourced code examples

| Type | Research Contribution | Research Advancement |
| --- | --- | --- |
| **Prevalence**: Empirical Evidence from Stack Overflow | Evidence of the Prevalence of Vulnerable C++ Code in Stack Overflow. We analyzed C++ code snippets contained in answers posted on Stack Overflow and identified the vulnerabilities that they contain. | Reusability of code/software is important to support modern day rapid software development [1] [2] [3] [4] [5] [6]. Significant research efforts are devoted to produce tools and techniques that can be used to produce live software documentation from Stack Overflow [19], or to detect/synthesize high quality posts in Stack Overflow to offer answers to a question [20], [21], [22], [23], [24]. Such tools and documentation from forums are important due to the shortcomings in software official documentation [25]. Our study recommends that such techniques can be further improved by adding security checks (e.g., for C++ code snippets) into the overall documentation or tool development process. |
| **Propagation**: Empirical Evidence from Stack Overflow and GitHub | Evidence of the propagation of Vulnerable C++ Code snippets from Stack Overflow to GitHub Repositories. We tracked all the vulnerable C++ code snippets found on Stack Overflow to their reusing projects on GitHub. We conducted a survey of GitHub developers who copied Vulnerable Code from Stack Overflow to their GitHub repositories. | One of the challenges in software is the reuse of vulnerable codes to accelerate the development of a software product, which ultimately leads to a decrease in software quality [8], [9]. In particular, research shows that developers in GitHub frequently visits developer forums like Stack Overflow to complete their coding tasks [26], and that common misuse patterns in code can be shared between the two sites frequently [27]. Our study complements the existing research by also showing that such phenomenon also exists for C++ code snippets, one of the most used languages in mission-critical systems. |

The 99 identified vulnerable code snippets are used in 2589 GitHub files. The most common vulnerability propagated from Stack Overflow to GitHub is CWE-150 (Improper neutralization of space/meta/control).

To assist developers in reusing code from Stack Overflow safely, we developed a browser extension that allows check-ing for vulnerabilities in code snippets when they see vul-nerable code snippets on Stack Overflow.

**Replication Package.** The corresponding data, survey mate-rials, and source code are shared in our GitHub and Zenodo repositories [17], [18].

**Paper Organization.** Section 2 provides background information about code reuse and discusses the related literature. Section 3 introduces our research questions, data collection, and data processing. Sections 4 and 5 discuss the obtained results, while Section 6 presents the results of a survey of GitHub developers and a browser plug-in that we de-veloped based on the survey findings. Section 7 discusses threats to the validity. Section 8 concludes the paper.

## 2 BACKGROUND AND RELATED WORK

In this section, we provide background information about security vulnerabilities and review the related literature.

### 2.1 CWE (Common Weakness Enumeration)

CWE is a community-developed list of common software security weaknesses. It serves as a common reference, a measuring stick for software security tools, and as a base-line for weakness identification, mitigation, and prevention efforts. It is regarded as an universal online dictionary of weaknesses that have been found in computer software. The purpose of CWE is to facilitate the effective use of tools that can identify, find and resolve bugs, vulnerabilities and exposures, in computer software before the programs are distributed to the public.

### 2.2 Reusing of Code Shared on Stack Overflow

Stack Overflow is regarded as the most popular question and answer website for software developers [16]. Software developers benefit from Stack Overflow posts, while pro-gramming [8], [12], [13], [28], [29], and read about the technologies and tools needed for development [30], [31], [32]. Thus, research on Stack Overflow is of high importance in software community.

Developers create and maintain software by standing on the shoulders of others [33]; they reuse components and libraries, and mine the Web for information that can help them in their tasks [34]. For help with their code, developers often turn to programming question and answer (Q&A) communities, most visible of which is Stack Overflow [35], [36]. Xia et al. [11] show that a large number of open source systems reuse outdated third-party libraries, which can lead to harmful effects on the software, because they may introduce security flaws in the software. Abdalkareem et al. [1] examined F-Droid repositories, and identified clones between Stack Overflow posts and Android apps. They observed that copied code from Stack Overflow posts can have an adverse effect on the quality of applications. Yang et al. [10] analyzed 909k non-fork Python projects hosted on GitHub, which contain 290M function definitions, and 1.9M Python code snippets reused from Stack Overflow. They performed a quantitative analysis of block-level code cloning intra and inter Stack Overflow and GitHub. Nishi et al. [37] studied code duplication between two popular sources of software development information: Stack Over-flow and software development tutorials, to understand the evolution of duplicated information overtime. An et al. [13] investigated clones between 399 Android apps and Stack Overflow posts. They found 1,226 code snippets which were reused from 68 Android apps. This reused of code snippets resulted in 1,279 cases of potential license violations. Baltes et al. [38] surveyed Stack Overflow users to understand the usage and attribution of code snippets in Stack Overflow. Among 122 respondents, 79% reported that they copied code from Stack Overflow no later than a month ago, and 39% not later than a week ago. Half of them (49%) copied the code without attributing the original Stack Overflow post, while the others added a source code comment with a link to the original Stack Overflow post.

Our study differs from the above studies in the following three aspects: 1) Previous studies have not investigated C++ programming language vulnerabilities in Stack Overflow. 2) Previous studies have not examined the relationship between Stack Overflow C++ posts and open source GitHub projects as well as migration of existing vulnerabilities through the reuse of Stack Overflow code snippets in open source GitHub projects thoroughly. 3) In this study, we have also analyzed the evolution of vulnerabilities in C++ Stack Overflow posts. No previous work analyzed the evolution of vulnerabilities within 10 years in Stack Overflow and their migrations to real-world projects.

### 2.3 Security Challenges of Code in Stack Overflow

Several studies have reported the presence of insecure code in some highly up-voted and accepted answers on Stack Overflow [8], [39], [40]. Rahman [41] applied topic model-ing on security-related discussions in Stack Overflow. They observed that the security topics can be grouped into five categories: 1) web security, 2) access control, 3) implementa-tion specific, 4) mobile security, and 5) system security. Yang et al. [42] also applied topic modeling to security related discussions in Stack Overflow. They also found similar top-ics from five categories: 1) web security, 2) mobile security, 3) cryptography, 4) software security, and 5) system security. Zhang et al. [27] investigated the quality of Stack Overflow code snippets by examining the misuse of API calls. They reported that approximately 31% of their analysed code snippets possibly incorporate API misuses that could lead to failures, such as, crash, resource leakages, etc.

Fischer et al. [8] extracted Android security-related code snippets from Stack Overflow, and manually labeled a sub-set of the data as "secure" or "insecure". The labeled data allowed them to train a classifier to efficiently judge whether a code snippet is secure or not. Next, they searched for code clones of the studied snippets in 1.3 million Android applications. They report that 15.4% of the Android applica-tions contained Stack Overflow source code. Of the analyzed source code, 97.9% contained at least one insecure code block. Thus, their work does not overlap with our study, Moreover, they studied Java-based systems while we focus on C++ systems.

These previous studies did not investigate C++ code snippets. Yet, C++ is the fourth most popular programming language. The use of insecure code snippets has been linked to multiple software attacks in which user credentials, credit card numbers, and other private information were stolen [43]. C++ is reported to be prone to misuses (e.g., memory corruption bugs) that easily lead to vulnerable code and exploitable applications [44], [45], [46]. In addition, our research advances the state of the art as follows: 1) To the best of our knowledge, our work is the first to investi-gate C++ vulnerability migration from Stack Overflow to GitHub. 2) The majority of studies that investigated vulner-abilities in Stack Overflow studied a limited time period, while we analyze vulnerabilities evolution over a period of ten years.

### 2.4 Security issues in GitHub

Rahman et al. [47] detected seven types of security smells that are indicative of security weaknesses in IaC scripts and identified 21,201 occurrences of security smells that include 1326 occurrences of hard-coded passwords. Zahedi et al. [48] examined issue topics in GitHub repositories and found that only 3% of them were related to security. The majority of these security issues were cryptography issues. Pletea et al. [49] examined security-related discussions on GitHub, and report that they represent approximately 10% of all discussions on GitHub. They also report that security related discussions are often associated with negative emotions. Acar et al. [50] conducted an experiment with active GitHub users to examine the validity of convenience sampling during the recruitment of participants in security-related studies. They observed that neither the self-reported status of participants (i.e., as student or professional developers) nor the security background of the participants correlated with their capacity to complete security tasks successfully.

### 2.5 Developer Studies in Secure Software Engineering

Qualitative methods have been used in software engineer-ing to study development, maintenance, and evolution prac-tices in real world settings. Qualitative data is often used in complement to quantitative data to increase confidence in the results of empirical studies through triangulation. An et al. [13] who investigated license violations in code reused from Stack Overflow conducted a survey of the developers of apps in which the violations were found. They contacted 23 developers and received six answers. All the six developers who replied confirmed the reported license violation issues. Acar et al. [7] surveyed 295 developers and conducted a user study with 54 students and professional Android developers. They found that most developers used search engines and SO to address security issues. Uddin et al. [51] surveyed software developers to understand why they seek opinions about software libraries in Stack Over-flow. They found that developers seek opinions to learn about diverse software aspects, including security. Uddin et al. [51] found that developers consider code examples with reviews (positive/negative sentiment) together to de-termine the quality of the provided code examples, and to them such shared knowledge is important due mainly to the shortcomings in official software documentation. In fact, previous surveys of developers in Industrial contexts (e.g., at IBM [25], Microsoft [52]) confirmed that developers find the official software documentation to be often incomplete, ambiguous and incorrect. These studies thus show that the shared code examples in Stack Overflow can be useful and important for developers while completing their development tasks. Therefore, the presence of bugs and security vulnerabilities in the shared code examples can introduce critical failure into their codebase.

TABLE 2: Comparison between our study and prior studies

| Theme | Our Study | Prior Study | Comparison |
|---|---|---|---|
| Reusability of C++ Posts in Stack Overflow | In this study we investigated the reusability of C++ code snippets from Stack Overflow answer posts to GitHub projects. | Reusability in Stack Overflow contain copying code in other open sources application, license violation in Stack Overflow and use in open source projects such as GitHub repositories [8] , [29], [34], [35]. The studies [10], [11], [12], [13] examined reusability in Java and Android application, [1], [53] in Python, [37], in third party code, [54] in IDE, [55] in API documentation, [56] in Php. | This paper examines the reusability of C++ code snippets from Stack Overflow to GitHub projects. No previous work in the literature have done a similar study. We study the prevalence of vulnerable C++ code snippets in Stack Overflow and their migration to GitHub projects. |
| Security of C++ Posts in Stack Overflow | Software security issues are broad and, at the same time, extremely difficult to detect; specially for C++ programming language. We analyze C++ code snippets in Stack Overflow answer posts. | Studies on Java Script [8], [39] and Android application [7], [40], [57] in Java and in Python [32] have shown the existence of security vulnerabilities in Stack Overflow code snippets reused in applications. | We carefully scrutinized each studied code snippet for security vulnerabilities and expressed each vulnerability with a CWE vulnerability label. No previous work has conducted such manual analysis and classification. |
| Security in GitHub Repositories | This study examined all GitHub projects that reused vulnerable C++ code from Stack Overflow over a period of ten years. | Previous studies on security in GitHub projects examined the following themes: secure coding, sentiment analysis, security issues [47], [48], [49], [50]. | No prior study specifically focused on vulnerable C++ code snippets that migrated from Stack Overflow to GitHub projects. |

Our findings in the paper show that the shared C++ code examples do contain critical security vulnerabilities. More worryingly, we observed that those vulnerable code snippets were reused in thousands of GitHub repositories. To inform the GitHub developers of those vulnerabilities, we conducted a small survey following principles and designs similar to the above research. Our proposed security fixes were accepted by some of those repositories.

## 3 RESEARCH QUESTIONS AND DATA COLLECTION

This section presents our research questions and describes our data collection and processing approach.

### 3.1 Research Questions

We explore the following Research Questions (RQs):

**RQ1: How prevalent are C++ vulnerabilities in Stack Overflow code snippets?**
Previous work on other programming languages revealed the existence of vulnerable code in Stack Overflow [7] [58]. To understand the existence and distribution of insecure C++ codes in Stack Overflow, we analyzed C++ answer posts throughout the ten years of Stack Overflow history.

**RQ2: How are the vulnerable C++ code examples shared on Stack Overflow reused in GitHub repositories?**
Knowledge sharing through code reuse routinely occurs between Stack Overflow and GitHub. The effect of vulnerable code migration to GitHub projects has not been investigated in details. Detected C++ vulnerable Stack Overflow code snippets might have migrated to GitHub and ended up deployed in the field. This research question aims to examine the extent of this phenomenon.

### 3.2 Data collection

In this section, we describe the data collection and analy-sis approaches that we used to answer our two research questions. Figure 1 shows a general overview of our data processing approach.

To study Stack Overflow posts evolution and their rela-tion with GitHub, SOTorrent data set version 2018-09-23 was used. In SOTorrent, each Stack Overflow post (question or answer) is identified by a Stack Overflow ID. All modifi-cations to the posts during the past ten years (2008-2018) are stored in the data set. Figure 2 shows the connections between Stack Overflow posts and their histories in the SOTorrent data set. Each post in the SOTorrent dataset may contain a url to a GitHub repository, if the url of the post is found in the GitHub repository. However, if a post is referenced in a GitHub repository, it is not clear which history record of the post in SOTorrent the GitHub reference belongs to. The SOTorrent data set that we used provides access to the version history of Stack Overflow content for ten years. In total, there are 41,472,536 posts in 2018. These question and answer posts were edited multiple times by Stack Overflow users. This resulted in 109,385,095 post versions. A Stack Overflow post may contain one or more code snippets that are tagged by markups. That means the textual and code contents in a post are separated based on markups (e.g., similar to < code >< =code > tag in Stack Overflow). For more detailed description on how SOTorrent is built please refer to [15]. In total 206,560,269 code and text blocks are identified in SOTorrent. Only 6,039,434 posts have identified links to software projects, where 3,861,573 had links to public GitHub repositories.

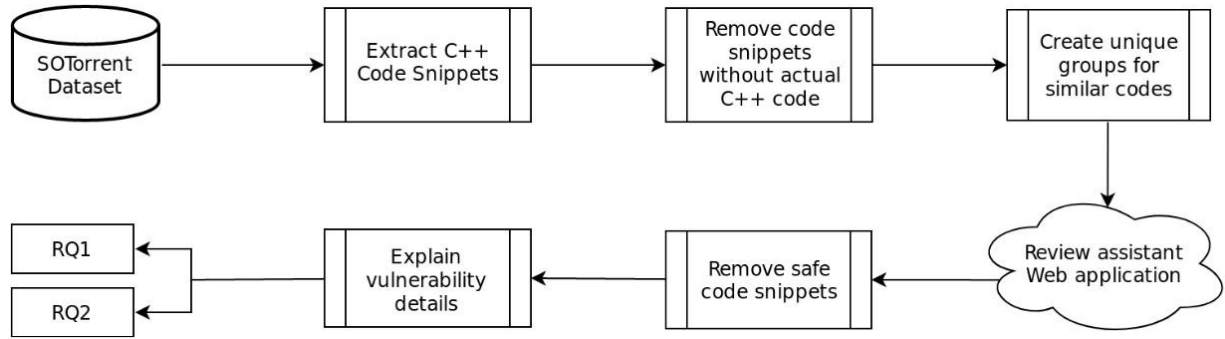

Fig. 1: Overview of our data processing approach.

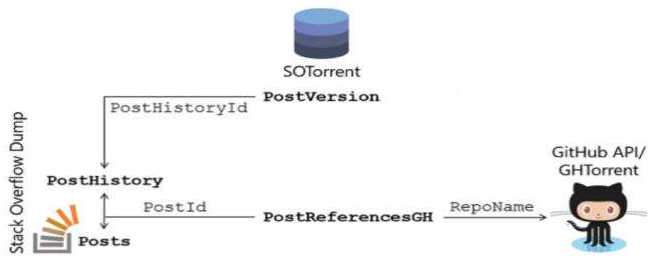

Fig. 2: Connection of SOTorrent table to other resources (simplified) [16]

### 3.3 Data Preprocessing

In our SOTorrent dataset, there were 583,415 questions tagged as C++ (out of 16,389,567 questions). These questions have 1,074,990 answers. Some answers were edited by users one or more times. The total number of answers is 1,738,346 if we also include the modified versions of answers.

Since our study aimed to analyze code snippets that migrated to GitHub, we removed answers that did not contain a code snippet. Not all the code blocks in a post may contain a valid code snippet. For example, some posts simply contain code markups, but we only found textual content inside those code markups. The total number of answer posts with one or more code blocks in our data set is 1,032,696.

For each answer in Stack Overflow with a C++ code snippet we take its url from Stack Overflow. We then check the url in the SOTorrent dataset, by consulting the table PostReferenceGH of SOTorrent data. The table tells us the list of GitHub source code files where the answer url is found, i.e., the code snippet may have been migrated to the GitHub source file. Overall, the 1,770 answers in Stack Overflow were found a total of 14,779 times in GitHub, i.e., they were referred to the various GitHub source code files a total of 14,779 times.

Since a vulnerability could exist in the code snippet of any version of an answer (i.e., including modified answers), and a developer could have reused that vulnerable code snippet into a GitHub file, we had to analyze all the code snippets extracted from all the versions of all the answers. In total, we identified 121,892 code snippets that could have migrated from Stack Overflow to a GitHub project.

### 3.4 Data cleaning

Not all code snippets in SOTorrent were actually C++ codes. Figure 3 shows a tagged code snippet that is supposed to be in C++, but which is actually a list of regular expressions for build files. Other examples of pseudo codes or plain texts tagged as code snippets could be found. Therefore, we needed a tool capable of identifying code snippets written in C++.

```
Add a list of your source files:
SOURCES = $(wildcard $(SRC)/*.cpp)
and a list of corresponding object files:
OBJS = $(addprefix $(TGT)/, $(notdir $(SOURCES:.cpp=.o)))
and the target executable:
$(TGT)/myapp: $(OBJS)
    $(CXX) $(LDFLAGS) $(OBJS) -o $@
```

Fig. 3: Example of code snippet with no real c++ code, but only configuration of 'makefile' (Answer 13109884)

Syntaxnet is a natural language parsing tool developed by Google research. Algorithmia has trained the Syntaxnet tool on a large number of programming language code blocks [59], such as C++, Java, etc. In this paper, we used this version of Syntaxnet trained by Algorithmia, to detect valid C++ code snippets. The Syntaxnet model takes as input a code snippet and outputs a confidence score (between 0 and 100) for a set of programming languages. For example, for a given code snippet in our dataset, the model can provide the following output: [(C++,95.5%), (Java, 32.3%), (PHP, 12.3%), (Perl, 2.8%)]. This means that the model has 95.5% confidence that the input code snippet is using C++ programming language, 32.3% confidence that the code snippet is written in Java. We assign a code snippet as a valid C++ code snippet, if the Syntaxnet model has the highest confidence score for C++ for the given code snippet (out of all the considered programming languages). At the end, among 121,892 possible code snippets extracted from Stack Overflow answers, only 72,483 code snippets were reported to be C++ code snippets by Syntaxnet. They came from 1,325 answers.

In SOTorrent, each change in a question or answer in Stack Overflow is stored as a set of records, but links to GitHub projects are provided only for the ID of answers or questions without direct indication of which version actually migrated. However, to investigate the migration of vulnerable code snippets to GitHub, we had to precisely identify the version of the post that migrated. Following previous research that focused on code examples shared only in answers while producing API documentation [19], in this study, we limited our analysis to answers with links to GitHub projects, because those code examples are more likely to offer solution to coding tasks.

of the 72,483 code snippets to GitHub, we extracted all the GitHub links contained in all the answers from which the 72,483 code snippets originated. The 2056 code snippets are extracted from 1,325 answers. An answer can be edited mul-tiple times, some time creating a slightly different version of a code snippet. We considered each such version as a distinct code snippet. This is why we have on average 1.55 code snippets per answer. In Figure 4, we show the popularity of the 1,325 posts based on two commonly used popularity metrics in the literature: number of scores (upvote - down-vote) and number of view counts. On average the answers have scores more than 82 and view counts more than 95K. Therefore, the answers and the shared code examples are very popular in Stack Overflow.

## 4 PREVALENCE OF C++ VULNERABILITIES IN STACK OVERFLOW CODE EXAMPLES (RQ1)

### 4.1 Approach

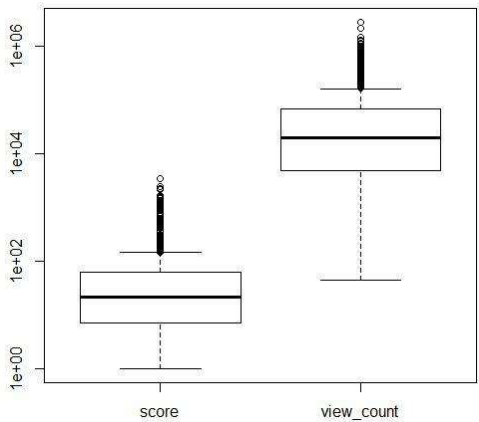

Fig. 4: Popularity of views and scores (Log 10 scale) of 1,325 answers of 2,056 code snippets groups

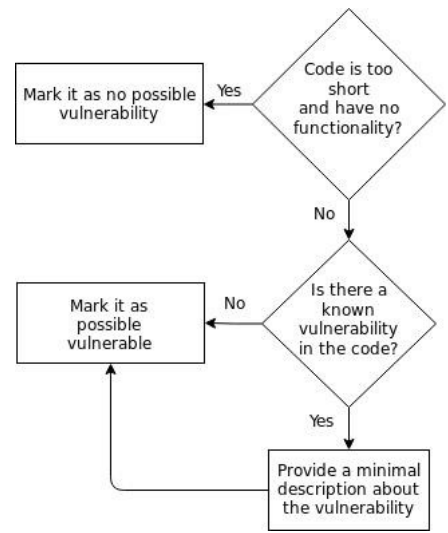

Fig. 5: Flowchart of code reviewing in first step

The SOTorrent data set that we used stores all the answers and questions of Stack Overflow with their history for ten years. Posts are broken into text and code and stored sepa-rately. Any change to the post within the period results in a replication of all the details of the answer or question in the corresponding tables that store the details of the post. Codes are stored as code blocks and as stated, any modification to the text of the answer or question results in storing replications of codes in code blocks. Therefore, another im-portant study design decision that we made concerns code duplication. We observed multiple cases of code duplication in our initial list of 72,483 code snippets extracted from Stack Overflow answers. We therefore applied SourcererCC [60] on the code snippets to group similar code snippets. We use Type-1 matches in SourcererCC to get exact clones. The input to SourcererCC was the list of all 72,483 code snippets. The output was a list of 2,056 clone groups. Each of the 72,483 code snippets is included in one of the 2,056 clone groups. Thus, our analysis of code snippets from the 2,056 clone groups is representative of the entire dataset (i.e., the 72,483 code snippets). The analysis of any of the code snippets from a group will give us the same information. Therefore, we randomly picked one code snippet from each group for our manual review analysis in Section 4.1.

To ensure that we did not miss any migration of each

In order to make the review process more efficient and systematic we created a web application having a simple interface with language-specific syntax highlighting. The web-based review application could mark code snippets as vulnerable, assign one or more CWE tags for each code snippet and view all similar code of a same answer at once. The review process had two stages of manual inspection.

Three of the main reviewers are from the Software Engi-neering research group at Shiraz University and contributed as authors (first, third and sixth authors). The first and third author had mastered Software Security in C++ by taking a graduate course on the subject under a professor of the area (second author). The sixth author was working as a professional penetration tester and mastered the official documentations on the subject. The other participants were volunteers of the same research group and contributed without any financial incentives. The collaborative nature of the large resource group enables the students to contribute to the research outcome of each other, which also serves as a learning experience for each student. As such, no direct incentive was involved. In the first round of reviews, the three experienced master students in C++ security reviewed 2,056 unique code snippets as noted before.

All code reviewers in this study passed the secure coding practical graduate course with C++ CERT [61] book as a reference over a period of one semester. They have all completed two practical and theoretical exams as a partial completion of their assessment in the graduate course of Software Security at Shiraz University.

At the first step of the manual inspection process, the goal was to reduce the size of the data-set without losing accuracy. Thus, all code snippets that were certainly not vulnerable were removed. Code snippets that were very short or did not have a specific functionality were removed. If a vulnerability within a code snippet was noticed within the first round of review, they would write a short description explaining why they thought the code snippet might be vulnerable and would add an appropriate CWE for the code snippet. During the review process, reviewers were directly in contact with each other and solved their disagreements through discussions. Figure 5 depicts the workflow and how our three reviewers inspected the code snippets and flagged them as vulnerable or not. After this first stage of thorough code review, 498 possible vulnerable code snippets were detected.

A second round of review was conducted using a set of guidelines established for the task. In order to find vulnerabilities in the studied code snippets, the reviewers needed to gain a good understanding of the code snippets and their evolution. Based on knowledge obtained from the first round of review and reading the main Software Security references [61], we established the following set of guidelines, with the aim to find as many vulnerabilities in the code snippets as possible.

1) **Read the question corresponding to the answer con-taining the probable vulnerable code snippet**: To have a better understanding of the reasons why developers shared the code snippet on Stack Overflow.
2) **Read the last version of the answer, its description, and analyze the evolution of the code over time**: To determine whether the vulnerability has been fixed or evolved within the various versions.
3) **Read the comments of the answers**: To determine if the vulnerability has been reported through the comments of the post. As an example, in Figure 6, 1st and 2nd comments indicated a vulnerability, and 3rd and 4th comments indicated a deprecated answer.
4) **Look for deprecated or dangerous functions in code snippet**: For example, the 'rand()' function is obso-leted since C++11 [62] and it is not recommended for random-number generation and cryptographic opera-tions.
5) **Check the arguments passed to the functions in the code snippet**: Types of arguments and their values are very important. For example, an out-of-bound large un-signed integer passed to a function that accepts signed integers may interpret the value as a negative number which would result in an undefined behaviour or a program crash.
6) **Check function usages based on official documentations**: For referencing and proper documentation of found vulnerabilities, official documentations were extensively used throughout the review process. For example, the 'malloc' function in C++ is used to allocate memory. Therefore, the documentation recommends to check its return value upon initialization to avoid problems, such as 'null dereferencing', which then create critical security vulnerabilities. Indeed, we found shared code snippets where the return value of malloc was not checked (e.g., see listing 2 in Section 4.2).
7) **Look for logical vulnerabilities in the code snippets** Usually, security is not the first priority of answerers in Stack Overflow. Answerers focus more on functionality than security. Therefore, the shared code snippet may miss obvious flaws that can introduce a critical vulnerability. For example, a shared code snippet may show how to read a vector in C++ without showing how to initialize it properly. Without proper bounds checking, this shared code snippet will introduce an index out of bounds problem in the reused code snippet. Indeed, we found such vulnerable code snippets in Stack Overflow. For example, see listing 5 in Section 4.2, where the goal is to read a vector, but no bounds checking is performed. Using a larger value than index bound can happen either by a programming mistake or could be the doing of an attacker.

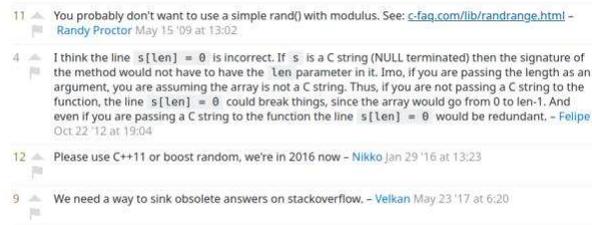

Fig. 6: Comments to vulnerable code in Answer (ID 440240)

After the second round of review, the identified vulnerable code snippets were confirmed and tagged based on CWE tags. One or multiple CWE tag(s) were assigned to each code snippet. These tags allowed us to track the evolution of the security of the code snippets throughout the evolution of Stack Overflow. We computed the Fleiss' Kappa [63] agreement among the three reviewers before discussions and obtained 0.26, which is a 'fair' agreement. Because the level of agreement between participants was only 'fair', the second author who is professor in software security organised a group discussion with 12 graduate students who participated in the manual evaluation to discuss each case and finalize the results of the first round of review, using majority votes. The output of the manual analysis is a list of vulnerable C++ code snippets found in Stack Overflow posts, with each vulnerability tracked to a CWE ID. The process took 868 hours.

TABLE 3: The different types of CWE C++ vulnerabilities and their frequency as we observed in our dataset of Stack Overflow Answers. Each tick in X-axis denotes the last one/two letters of a year, e.g., 8 for 2008 to 16 for 2016.

| CWE | Title and Description | Frequency by Year |
|---|---|---|
| 1006 | **Bad Coding Practices** These weaknesses are deemed to cause exploitation's that are not vulnerable by self but indicate that the application is not developed carefully. | 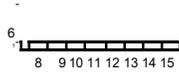 |
| 754 | **Improper Check for Unusual or Exceptional Conditions** This vulnerability occurs based on the assumption that events or specific circumstances never happen, such as low memory conditions, lack of access to resources. | 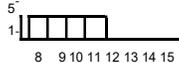 |
| 20 | **Improper input validation:** When software does not validate input properly, an attacker is able to craft the input in a form that is not expected by the rest of the application. | 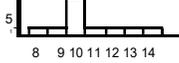 |
| 252 | **Unchecked return value** The return value is not checked by a method or function, which may create an unexpected state. | 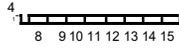 |
| 1019 | **Validate input Weaknesses** to a degradation of the quality of data flow in a system. | 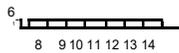 |
| 477 | **Use of obsolete function** The code uses deprecated or obsolete functions, which suggests that the code has not been actively reviewed or maintained. | 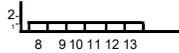 |
| 789 | **Uncontrolled memory allocation** Memory is allocated based on invalid size and allowing arbitrary amounts of memory to be allocated. | 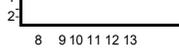 |
| 158 | **Improper neutralization of null byte or null character** The input is received from a upstream component, but it does not neutralize or incorrectly neutralizes when null bytes are sent to a downstream component. | 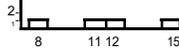 |
| 134 | **Use of externally controlled format string** Have been used a function that accepts a format string as an argument, but the format string originates from an external source. | 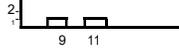 |
| 476 | **Null pointer deference** A NULL pointer dereference occurs when dereference a pointer that it expects to be valid, but is NULL, typically causing a crash or exit. | 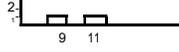 |
| 121 | **Stack base buffer overflow** The situation is where the buffer is rewritten in the stack (like, a local variable or, rarely, a parameter to a function). | 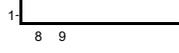 |
| 415 | **Double free** Called free() twice on the same memory address, potentially leading to modification of unexpected memory locations. | 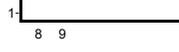 |
| 676 | **Use of potentially dangerous function** Invoked a potentially dangerous function that could introduce a vulnerability if it is used incorrectly. | 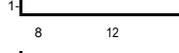 |
| 628 | **Function call with incorrect specific arguments** The product calls a function, procedure, or routine with arguments that are not correctly specified, leading to always-incorrect behavior and resultant weaknesses. | 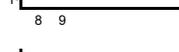 |
| 686 | **Function call with incorrect argument type** A function, procedure, or procedure is called up with arguments that are not properly specified, resulting in always mistaken behavior and resulting weaknesses. | 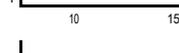 |
| 835 | **loop with unreachable exit condition** The program contains an iteration or loop with an exit condition that cannot be reached, i.e., an infinite loop. | 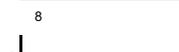 |
| 369 | **Divide by zero** Typically occurs when an unexpected value is provided to the product, or an error occurs that is not properly detected. | 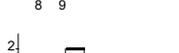 |
| 116 | **Improper encoding or escaping of output** A structured message is prepared to communicate with another component, but encoding or escaping of the data is either missing or done incorrectly. | 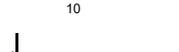 |
| 232 | **Improper handling of undefined values** Does not handled or incorrectly handled when a value is not defined or supported for the associated parameter, field, or argument name. | 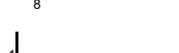 |
| 758 | **Reliance on Undefined unspecific or implementation defined behavior~** Used an API function, data structure, in a way that relies on properties that are not always guaranteed to hold for that entity. | 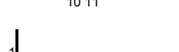 |
| 120 | **Classic buffer overflow** Been copied an input buffer to an output buffer without verifying that the size of the input buffer is less than the size of the output buffer. | 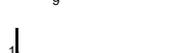 |
| 682 | **Incorrect calculation** Perform a calculation that generates incorrect or unintended results that are later used in security-critical decisions or resource management. | 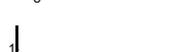 |
| 190 | **Integer overflow or wraparound** Perform a calculation that can produce an integer overflow or wraparound, when the calculation is used for resource management or execution control. | 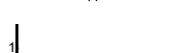 |
| 413 | **Improper resource locking** The software does not lock or does not correctly lock a resource when the software must have exclusive access to the resource. | 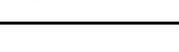 |

Continuation of Table 3: The different types of CWE C++ vulnerabilities and their frequency as we observed in our dataset of Stack Overflow Answers. Each tick in X-axis denotes the last one/two letters of a year, e.g., 8 for 2008 to 16 for 2016.

| CWE | Title and Description | Frequency by Year |
|---|---|---|
| 710 | **Improper adherence to coding standards**<br>Not followed certain coding rules for development, which can lead to resultant weaknesses. | 1 – 11 |
| 150 | **Improper neutralization of escape, meta, or control sequence**<br>The software receives input from an upstream component, but it does not neutralize or incorrectly neutralizes special elements when they are sent to a downstream component. | 1 – 11 |
| 78 | **OS command injection:**<br>Constructs all or part of an OS command using externally-influenced input from an upstream component, that could modify the intended OS command when it is sent to a downstream component. | 1 – 12 |
| 426 | **Untrusted search path**<br>The application searches for critical resources using an externally-supplied search path that can point to resources that are not under the application's direct control. | 1 – 10 |
| 193 | **Off by one error**<br>A product calculates or uses an incorrect maximum or minimum value that is 1 more, or 1 less, than the correct value. | 1 – 8 |
| 131 | **Incorrect calculation of buffer size**<br>Does not correctly calculate the size to be used when allocating a buffer, which could lead to a buffer overflow. | 1 – 13 |
| 125 | **Out-of-bounds Read**<br>The software reads data past the end, or before the beginning, of the intended buffer. | 1 – 16 |

## 4.2 Results

In Table 3, we summarize the list of CWE vulnerabilities that were found in the code snippets during our manual analysis. The first column ('CWE') in Table 3 provides the ID of the vulnerability from the CWE database. The second column ('Title and Description') presents a brief description of the vulnerability. Please see [64] for a complete descrip-tion of each CWE vulnerability. A total of 31 different CWE vulnerabilities were found in 99 vulnerable code snippets. We present those vulnerabilities in Table 3 based on their occurrence frequency in the 99 code snippets, i.e., the most frequent vulnerability is placed at the top. The vulnerability most frequently observed in our manual analysis was 'Bad Coding Practices' (CWE ID 1006), followed by 'Improper check for unusual or exceptional conditions' (CWE ID 754), and 'Improper code validation' (CWE ID 20). Indeed, im-proper or insufficient checks of inputs could create many critical C++ vulnerabilities. For example, the 'Improper code validation' vulnerability occurs when, for example, a buffer in C++ is not checked for size before providing input. This lack of checking then can create critical security attacks, such as Buffer Overflow, which is frequently exploited by hackers to gain unauthorized access to a system or to create Denial of Service (DoS) attack against a system. The last column ('Frequency by Year') in Table 3 shows the distribution of the CWE vulnerability in our manually ana-lyzed code snippets. We show the distribution by year. For example, the vulnerability 'Improper input validation' was frequently observed in all the years between 2008 and 2014. The SOTorrent dataset was created from Stack Overflow. Stack Overflow was created in 2008. Some vulnerabilities were not observed in older code snippets, but are found in the newer code snippets. For example, the 'Out-of-bounds Read' vulnerability (CWE ID 125) only started to show in the shared code snippets around 2016. In Stack Overflow, older posts can be as popular as a new one and thus older code examples can still be reused. This is especially true for C/C++-based systems, where legacy APIs are widely used in mission critical systems. Overall, the diverse distributions of the vulnerabilities across the years shows the challenge developers can face while trying to reuse those code snip-pets, especially when they are not security experts.

In the following, we present some examples of vulnera-bilities found in the inspected code snippets.

Listing 1: Generate random string in C++ - Answer id 440240 in Stack Overflow, shows vulnerability due to use rand function with incorrect using method, (CWE-1006, CWE-477, CWE-193, CWE-754)

```cpp
void alphanum [ gen random (char *s, const int len) ] f
    static const char alphanum[] =
        "0123456789"
        "ABCDEFGHIJKLMNOPQRSTUVWXYZ"
        "abcdefghijklmnopqrstuvwxyz";
    for ( int i = 0; i < len; ++i) f
        s[ i ] = alphanum[ rand()% ( sizeof(alphanum))];
    g
    s[len] = 0;
g
```

The code snippet of answer 440240 shown in Listing 1 can be dangerous. Functions with count parameters like 'len' should take into account the terminating 'NULL' as an extra character. But this function actually writes into the character 'len+1' when executing s[len] = 0. That is CWE-193:Off-by-one-error vulnerability [65]. A vulnerability that may lead to unpredictable behaviour, memory corruption and application crash. The function only works if less than permitted length is used. For example, Line 's[i] = alphanum[rand() % (sizeof(alphanum)]' is faulty since size of 'alphanum' is '63', where the last character in the string indexed $62^{nd}$ is 'NULL'. Therefore, once in a while a NULL may be included in the generated 'random' string. This vulnerability can be categorized as 'CWE-754: Improper check for unusual or exceptional conditions' [66], where an improper number may be used as a return of a function leading to a crash or other unintended behaviours. Another appropriate category is 'CWE-1006: Bad coding practices'. Stated differently, a generated random string with this algorithm may include 'NULL' in the middle of string.

Moreover, 'rand()' is an obsolete function in C and C++. So another vulnerability category is 'CWE-477: Use of obsolete function' a major degradation in software quality. Another vulnerability exists within the code since the developer did not use a random seed before calling the function. Thus, the generated random number is not 'random' at all. Moreover, 'rand() % mod' is not a good practice since it returns lower bits which are again not random [67].

Listing 2: Execute functor in given thread in QT - Answer id 21653558 in Stack Overflow, shows vulnerability due to use malloc function without checking return special condition, (CWE-1006, CWE-252, CWE-789, CWE-476)

```cpp
class FunctorCallEvent: public QMetaCallEvent {
public:
    template <typename Functor>
    FunctorCallEvent(Functor && fun, QObject * receiver) :
        QMetaCallEvent(new QtPrivate::QFunctorSlotObject<Functor,
            0, typename QtPrivate::List_Left<void, 0>::Value, void>
            (std :: forward<Functor>(fun)), receiver, 0,
            0,0,( void**) malloc (sizeof(void*)));
```

Another vulnerability is shown in Listing 2 of answer 21653558. The code snippet in this answer uses 'malloc' to allocate memory and passes its pointer to a function in QT library that requires a valid pointer. The malloc return pointer may be set to NULL in case of malloc failure. Thus, the return pointer from malloc must be checked even if the amount of memory requested is small [68]. In this example, the return value of malloc is not checked. This vulnerability is called CWE-252 [69]; Unchecked return value. In case of malloc failure, null pointer dereference occurs.

Listing 3: Execute command and get output - Answer id 478960 in Stack Overflow, Execute function in given thread in QT due to OS command injection because user input are involved, (CWE-78, CWE-1019)

```cpp
std :: string exec(const char* cmd) {
    std :: shared_ptr<FILE> pipe(popen ( cmd , "r"), pclose);
    if (! pipe) return "ERROR";
    char buffer[128];
    std :: string result = "";
    while (! feof (pipe.get () ) ) {
        if ( fgets (buffer , 128, pipe.get () ) != NULL)
            result += buffer;
    }
    return result ;
}
```

The function shown in Listing 3 is vulnerable to code injection (OS command injection) attacks since user inputs commands are inputted and not checked. In other words, any command with privilege level of the program can be executed without any errors or warnings.

Listing 4: Set the global LUA_PATH variable programmatically - Answer id 4156038 in Stack Overflow, shows vulnerability due to second arg in this function may contain multiple path separated by ";" (CWE-754, CWE-252, CWE-426)

```cpp
int setLuaPath( lua_State* L, const char* path ) {
    lua_getglobal( L, "package" );
    lua_getfield ( L, 1, "path" );
    std :: string cur_path = lua_tostring( L, 1);
    cur_path.append( ';' ) ;
    cur_path.append( path );
    lua_pop( L, 1 ) ;
    lua_pushstring( L, cur_path.c_str () ) ;
    lua_setfield ( L, 2, "path" );
    lua_pop( L, 1 ) ;
    return 0;
}
```

Listing 4 deals with system path programmatically, or different paths the program searches. The operation is dangerous and should be performed carefully. For example, 'path' in this function may contain multiple paths separated by ';'. For instance, '/usr/share/lua;/foo/bar/evil/path'. Having an untrusted search path within the paths produces the probability of arbitrary code execution with privilege of the program and redirection to a wrong file potentially triggering a crash. The vulnerability is called CWE-426: Untrusted search path. For more on this vulnerability, please refer to [70]. The search may lead to execution of programs, which in turn may lead to unusual or exceptional conditions; i.e., CWE-754: Improper check for unusual or exceptional conditions. Moreover, all the return values of the functions in the code snippet are not checked. Thus, the snippet also has CWE-252: Unchecked return values.

Listing 5: Set Byte vector to integer type - Answer id 41031865 in Stack Overflow, shows vulnerability due to out of bound read and the lack of checking the size of the variable, (CWE-20, CWE-125, CWE-1019)

```cpp
template<typename T>
static T get_from_vector(const std::vector<uint8_t>& vec,
        const size_t current_index )) {
    T result;
    uint8_t *ptr = (uint8_t *) &result;
    size_t idx = current_index + sizeof(T);
    while(idx > current_index)
        *ptr++ = vec[ idx];
    return result ;
}
```

In Answer 41031865 shown in Listing 5, 'current_index + sizeof(T)' can become larger than size of 'vec' due to CWE-1019: Validate inputs vulnerability. In addition, when index exceeds the limit, information leakage can occur or CWE-125; the vulnerability 'Out of bound read' is present.

Listing 6: Checks if string ends with .txt - Part of answer id 20447331 in Stack Overflow, all defined functions have vulnerability have fail if input string that contain a null value, (CWE-158, CWE-1019)

```cpp
bool ends_with (std :: string const &a, std :: string const &b) {
    auto len = b.length() ;
    auto pos = a.length()     len;
    if (pos < 0)
        return false ;
    auto pos_a = &a[pos];
    auto pos_b = &b[0];
    while (*pos_a)
        if (*pos_a++ != *pos_b++)
            return false ;
    return true;
}

bool ends_with_string (std :: string const& str, std :: string const&
        what) {
    return what.size() <= str.size ()
        && str.find(what, str . size ()     what.size()) != str . npos;
}
```

In answer (20447331) shown in Listing 6 on how to validate whether a file name ends with ".txt" or not, this answer includes code of functions and their benchmarks for six methods in the original code snippet in answer post. The vulnerability for other functions defined in the code snippet is exactly the same as the two vulnerable functions. However if filename in function includes a NULL character, all of above methods will fail. This is a common trick to bypass web application firewalls and file upload applications. For example, Validating 'shell.txtn0.php' will return True for all of above functions.

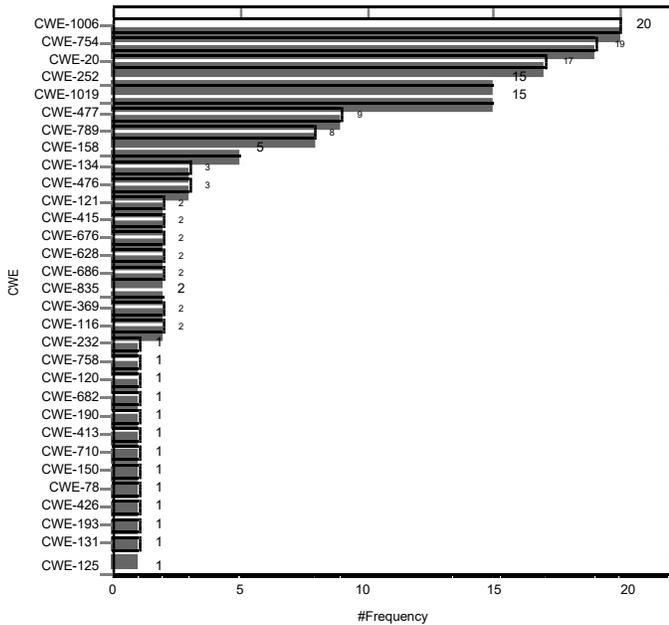

Fig. 7: Frequency of CWEs in code snippets

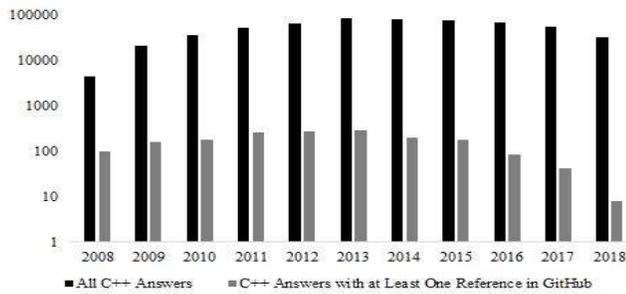

Fig. 8: Distribution of answers in C++ by year

From our manual reviews of the code snippets, we found 99 vulnerable code snippets residing in 69 answers. The frequency of CWE's in code snippets is presented in Figure 7. CWE-1006 (Bad coding practices) and CWE-754 (improper check for unusual or exceptional conditions) are the most frequent ones, followed by CWE-20 (improper input validation). Two of the top three vulnerabilities are related to the improper or lack of checking of inputs and conditions, i.e., developers who shared the code were either not aware of those potential vulnerabilities or they are not careful enough. Given that those shared code snippets are found in popular questions and answers in Stack Overflow, they were nevertheless reused by other developers.

The distribution of all C++ answers from 2008 to 2018 is shown in Figure 8. If one hypothesizes that Stack Overflow usage reflects the popularity of the programming language, C++ has been the most popular programming language in 2013, and its usage declined after that. By looking at the distribution of answers, we find that most answers were created in 2013 (as shown in Figure 8). The distribution of Stack Overflow answers linked to GitHub projects by year again shows that in 2013, C++ had the most migrations to GitHub projects (see Figure 8).

## 5 Propagation of C++ Vulnerable Code from Stack Overflow to GitHub (RQ2)

### 5.1 How frequently are the vulnerable code examples from Stack Overflow copied to GitHub? (RQ2.1)

To detect the vulnerable code snippets that migrated to GitHub projects, it may seem plausible to use clone detec-tion tools like SourcererCC [60]. However, the most effective clone detection tools work only for Java applications, e.g., Oreo [71]. The ones that can detect C++ clones only work at file or class level. For Java, SourcererCC can find cloned procedures but the same capability is not implemented for C++. The majority of vulnerable code segments that we found are functions or a part of a function. Therefore, we had to use some heuristics to search and find similar codes in linked GitHub projects. To find vulnerable clones, we searched for the signatures of the code snippets in Stack Overflow by looking at the sequences of keywords that can uniquely characterize them within GitHub projects. To derive our heuristics, we took inspiration from previous work [72], [73] and opted for a rule-base approach. We chose rules because unlike keyword-based searching, rules are less susceptible to false positive [72], [73]. For each code snippet, we selected an ordered sequence of sub-strings that can be used to determine the presence of the code snippet in the linked GitHub projects.

For example, to detect vulnerable GitHub projects that used 'rand()%( sizeof(alphanum))]' like Listing 1, we search for 'RAND()' sub-string and for versions that had 'rand()%( sizeof(alphanum)-1)]' we look for two consecutive sub-strings of 'RAND()' and ')-1)'.

The total count of GitHub files for the 69 vulnerable answers was 2,859 GitHub links. We present them in Table 5 along with the CWE definition. After executing our pro-posed tracing approach, we found 287 GitHub files that may contain the security flaws imported from Stack Overflow. After a careful manual review of the files, we found the tracing approach to be 78.72% accurate (see Table 4 below).

In the following, we compare the proposed approach with a BASELINE approach in which we have changed the criteria for choosing the keywords. In the BASELINE approach, we chose the keywords randomly without changing the number of words selected for each code snippet. In order to select random keywords, we first removed the comments of the code snippet and then tokenize the code snippet. We removed reserved keywords from the code snippets and only chose words with more than three characters as candidate keywords.

Table 4 presents the results obtained when comparing our proposed algorithm with the baseline method on 296 files. The columns 'Chosen-Keyword' and 'Random-Keyword' report the number false positive, true negative, false neg-ative, and true positive for respectively our proposed ap-proach and the BASELINE approach.

For each algorithm and for each of the code snippets found in the 69 answers in Stack Overflow, we executed the algorithm as follows. Using the SOTorrent database, we first collected the list of all GitHub files where a vulnerable code snippet might have migrated. If the SOTorrent database reported more than five such GitHub files, we randomly selected five GitHub files and ran our algorithm to deter-mine whether the vulnerable code snippet was reused in each of the GitHub files. If the SOTreent database reported less than five such GitHub files, we ran our algorithm on all those GitHub files. For each such GitHub file, the algorithm returns a '1' if it considers that the vulnerable code snippet is reused in the GitHub file i.e., the file is vulnerable due to reuse of the code snippet. It returns 0 otherwise i.e., the file is not vulnerable. Based on this setup, we computed the four metrics. True Positive = The algorithm considers a file vulnerable, the file is actually vulnerable. False Positive = The algorithm considers a file vulnerable, but the file is not actually vulnerable. True Negative = The algorithm does not consider a file vulnerable and the file is actually not vulnerable. False Negative = The algorithm does not consider a file vulnerable but the file is actually vulnerable.

TABLE 4: Comparison between our proposed approach and the BASELINE approach

|  | Chosen-Keyword | Random-Keyword |
| --- | --- | --- |
| False Positive | 34 | 60 |
| True Negative | 10 | 160 |
| False Negative | 29 | 41 |
| True Positive | 223 | 35 |
| Accuracy | 78.72% | 65.87% |

For each method, we calculate Recall, Precision, F1-Measure and Accuracy on all analyzed GitHub projects:

$$RECALL = \frac{TruePositive}{TruePositive + FalseNegative}$$

$$PRECISION = \frac{TruePositive}{TruePositive + FalsePositive}$$

$$F1\ Measure = 2 * \frac{Precision * Recall}{Precision + Recall}$$

$$ACCURACY = \frac{TruePositive + TrueNegative}{TruePositive + FalsePositive + FalsePositive + FalseNegative}$$

As shown on Figure 9, our proposed approach significantly outperforms the BASELINE approach. The most important weakness of the Random-Keyword is the use of keywords that do not indicate the existence of security vulnerabilities in GitHub code. Listing 7 presents the ordered sequence of sub-strings used to identify files having the vulnerability of answer ID 4156038. The random set of

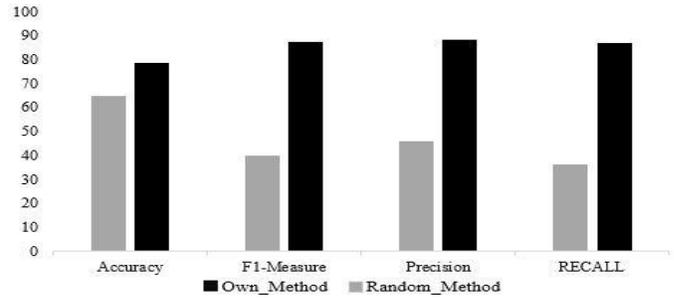

Fig. 9: Comparison between our proposed approach and the BASELINE approach

keywords selected by the BASELINE approach could not capture the vulnerable part of the code and its sequence, and therefore failed to track the vulnerable code snippet.

Listing 7: Selected keywords from answer id: 4156038

```
!"Path"
! lua_tostring(
```

More details about our tracing approach and its evaluation is available in our online appendix [17] [18].

## 5.2 How frequently are the copied vulnerable code examples fixed in the GitHub repositories? (RQ2.2)

We created a new web application to make the review process more efficient and systematic. With the review system, we assessed whether the vulnerable Stack Overflow code snippets that were migrated to GitHub were either fixed or still contained the vulnerability.
Among the 287 GitHub files that we checked, vulnerabilities were corrected in 34 files and the other 253 GitHub files still had vulnerabilities at the time of writing of this paper. For instance, as can be seen in Listing 8, the two CWE-789 [74] and CWE-252 [69] were corrected.

Listing 8: Part of code was Fixed and improved in GitHub File for answer id 2654860 in Stack Overflow

```
//Improve and adapted version of http://stackoverflow.com/a/
    2654860
void save_bmp(string filename, uchar4* ptr, const int width, const int height)
f
const int num_elems = width*height;
unsigned char* img = (unsgined char*)malloc(3* num_elems); int i =0;
 .....
g
```

As shown in listing 9, boundary was limited and mentioned in comments of source code in GitHub file and CWE-125 [75], CWE-category-1019 [76] and CWE-20 [77] were corrected.

Listing 9: Code was Fixed in GitHub File mentioned in comment for answer id 41031865 in Stack Overflow

```
//check if we can read sizeof(T) bytes starting the next index
check_lenght(vec.size() , sizeof (T), current index + 1); T result ;

auto* ptr = reinterpret_cast <uint8 t*>(&result);
```

TABLE 5: CWE's Detection in GitHub Repositories with chosen-keyword Algorithm

| GitHub Count | Confirm Count | CWE Title |
|---|---|---|
| 1539 | 4 | CWE-835-Loop with unreachable Exit condition |
| 703 | 37 | CWE-20-Improper input validation |
| 653 | 72 | CWE-754-Improper check for un-usual or exceptional condition |
| 324 | 187 | CWE-1006-Bad coding practice |
| 250 | 5 | CWE-158-Improper neutralization of null byte or null character |
| 212 | 2 | CWE-369-Divided by zero |
| 151 | 141 | CWE-150-Improper neutralization of escape, meta, or control sequence |
| 118 | 0 | CWE-628-Function call with incor-rectly specific argument |
| 89 | 14 | CWE-252-Unchecked return value |
| 73 | 2 | CWE-134-Use of externally con-trolled format string |
| 54 | 4 | CWE-476-Null pointer dereference |
| 53 | 4 | CWE-789-Uncontrolled memory al-location |
| 41 | 12 | CWE-477-Use of obsolete function |
| 20 | 1 | CWE-676-Use of potentially danger-ous function |
| 20 | 0 | CWE-232-Improper handling of un-defined values |
| 14 | 2 | CWE-121-Stack base buffer overflow |
| 7 | 0 | CWE-415-Double free |
| 5 | 1 | CWE-78-Improper neutralization of special elements used in an os com-mand |
| 5 | 0 | CWE-413-Improper resource locking |
| 5 | 5 | CWE-116-Improper encoding or es-caping of output |
| 5 | 0 | CWE-193-Off by one error |
| 3 | 3 | CWE-682-Incorrect calculation |
| 3 | 0 | CWE-686-Function call with incor-rect argument type |
| 3 | 0 | CWE-120-Buffer copy without checking size of input |
| 3 | 0 | CWE-131-Incorrect calculation of buffer size |
| 3 | 1 | CWE-710-Improper adherence to coding standard |
| 1 | 0 | CWE-426-Untrusted search path |

```
for ( size_t i = 0; i < sizeof(T); ++i)
f
    *ptr++ = vec[current index + sizeof(T)                i];
g
return result ;
g
```

## 6 DISCUSSIONS

In this section, we first describe a survey of GitHub developers who reused the C++ vulnerable code snippets in their GitHub repositories (see Section 6.1). We then describe a browser plug-in that we developed to warn such unsuspecting users of potential vulnerabilities in C++ code snippets in Stack Overflow (see Section 6.2).

### 6.1 Reaction from GitHub developers about the vulnerable code examples

To inform developers about the vulnerabilities found in their repositories, we developed a script to automatically report issues to the projects repositories. The script was fed with the results of our code review. We provided developers with the following information:

**Description:** The vulnerability in the code snippet is explicitly expressed.
**Example:** An attack scenario is provided to justify why the vulnerability is dangerous and how it may lead to exploitation.
**Mitigation Scenario:** The mitigation scenario is in-cluded to inform the developer on how to fix the vulnerability.
**Reference:** An authenticated reference is provided to show that the vulnerability was labelled with a CWE ID based on objective and factual judgements.

In addition, five questions related to the vulnerability were asked to the developers who responded to our fixes. Depending on the type of the question, we provided the developers with a plain text box or a Likert scale, to answer the question. Table 6 shows the survey questions and the type of answers that we recorded for each question.

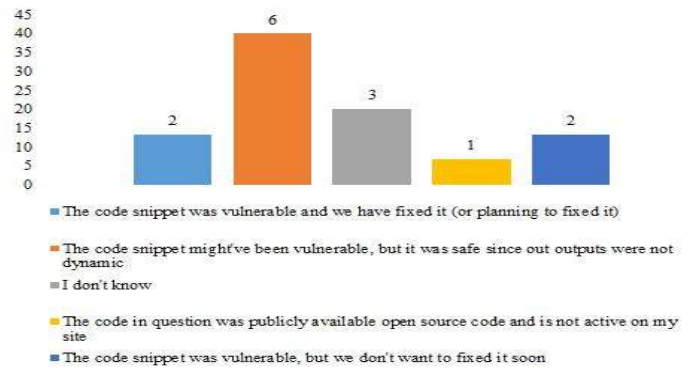

Fig. 10: GitHub developer opinion about the vulnerabilities

We received 15 responses from the 174 issues that were sent. Figure 10 breaks down the 15 responses based on respondents' opinions about the reported vulnerabilities. None of the respondents disputed the validity of the reported vulnerabilities, i.e., our assessment of vulnerabilities was reliable. However, only two out of the 15 respondents acknowledged that they fixed the vulnerability after our recommendation. Eight out of 15 opted to keep the original code as is, arguing that the underlying code base (where the vulnerable code snippet is reused) will not be exposed online and that the vulnerability won't therefore be exploitable. This is of course a dangerous assumption; given how any system component can be reused. Some of the respondents also personally thanked us for reporting the issues. Figure 11 shows an example response.

When we asked the respondents whether automatic tool supports to detect those vulnerabilities would be useful, 14 out of the 15 respondents agreed (7 agree + 7 strongly agree) that such a tool could help them (see Fig 12). When we further asked the respondents about the specifics of such automated tools (see Figure 13), 7 respondents asked for a browser plug-in, 7 requested a tool that informs devel-opers via automatically generated emails, 6 requested an offline tool, and 2 asked to be notified about vulnerabilities through an online repository of website. Therefore, the re-spondents preferred to be notified instantly when possible, e.g., through a browser plug-in that can warn them of the potential vulnerability in a code snippet during their visit in Stack Overflow.

TABLE 6: Questions asked in survey.

| NO | Question |
|---|---|
| 1 | **Which of the following situation for our issue was true?** (eight options) |
| 2 | **Please justify your choice above** (text box) |
| 3 | **What would be the best way to inform developers of potential vulnerabilities in code examples shared on Stack Overflow** (5-point Likert scale for each opinion) |
| 4 | **Do you have any other suggestions to design automated techniques to assist developers to handle security vulnerabilities while using code from online forums? Please write in a couple of sentences below** (text box) |
| 5 | **Could automated vulnerability analyses of code snippets in online forums be useful in your future development tasks?** 5-point Likert scale for each opinion |

Fig. 11: User response about security vulnerability in code

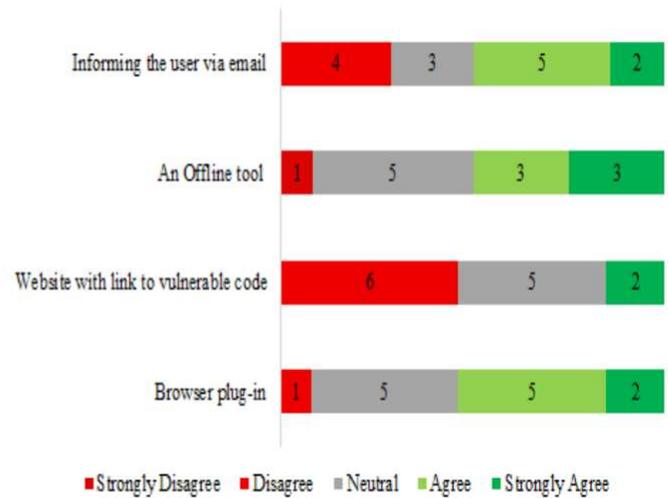

Fig. 13: User Opinion About How To Inform Developers of Potential Vulnerabilities in Code Examples

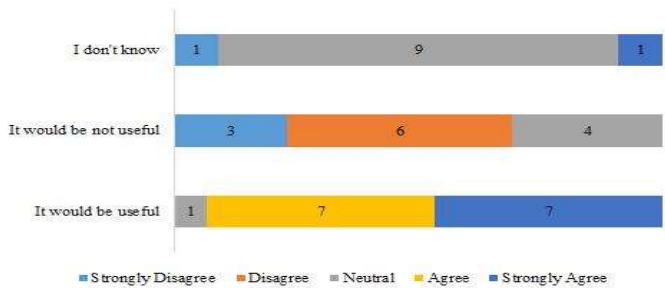

Fig. 12: User Opinion About Automated Vulnerability Analysis Being Useful for Future Development

### 6.2 A Browser Extension To Warn SO Users About C++ Vulnerability in Stack Overflow

To inform users about the existence of vulnerabilities in a code snippet posted on Stack Overflow, we have developed a browser extension. In Figure 14, we show a screenshot of our developed extension. The extension gets activated when a developer visits a Stack Overflow post. The extension consults our database of vulnerable C++ code snippets in Stack Overflow to determine whether the provided solution in the post is vulnerable. If the provided solution is indeed found vulnerable, the extension then shows a warning mes-sage to the developer with an explanation of why the code snippet is vulnerable (see Figure 14). The extension then rec-ommends non-vulnerable similar code snippets from other Stack Overflow posts, so that the developer can reuse those safe code snippets instead of the vulnerable code snippet.

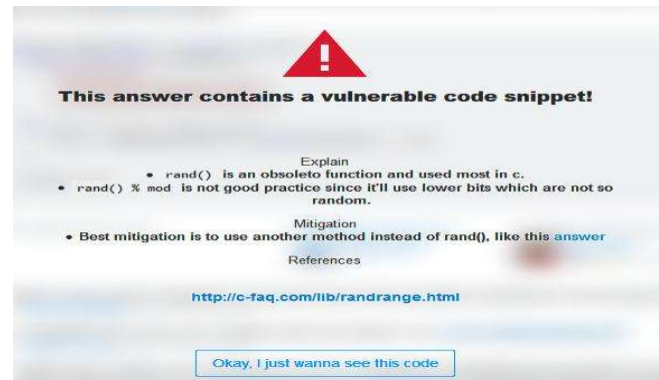

Fig. 14: Browser Extension for Code Snippet with AnswerId 440240 in Stack Overflow

Suppose that a developer needs to create a random alpha-numeric string in C++ for their task in their program. The developer searches in Stack Overflow for a possible solution. The search shows a question with ID 440133 as the top match. The importance of the question is determined in Stack Overflow based on how developers perceive the question. The asker of this question offered a bounty reward of 100 to the accepted answer. Consequently, the question received many answers. The accepted answer (ID 440240) has 263 scores (upvote - downvote) and it was viewed more than 174,000 times as of today. Therefore, a new developer looking for a solution for this task is expected to be con-vinced to use the solution provided in the answer. However, the provided solution has one of the security vulnerabilities presented in Section 4.2 (Listing 1). Therefore, the provided solution, if used as is, will introduce potential C++ secu-rity vulnerability in the developer's software. Our browser extension aims to prevent developers from reusing such vulnerable code snippets, as well as to recommend them of better alternatives, i.e., non-vulnerable code snippets in other Stack Overflow posts. As we recall from our survey of GitHub developers, such a browser extension was also desired by the survey respondents.

The design of the plug-in allows it to be programming language-neutral, i.e., the underlying architecture of the plug-in allows to be used to warn of security vulnerabilities for any given programming languages. The architecture is REST-based, i.e., this is a client-server model where the plug-in acts as the client and the security vulnerabilities database resides in a server. Therefore the database can be updated with new vulnerabilities information from any programming language. The plug-in only needs to know the ID of a Stack Overflow post and the code example provided in the post to consult with the server about whether the shared code example is vulnerable or not.

While the focus of this paper is to study the prevalence and propagation of vulnerabilities of C++ code examples shared in Stack Overflow, the plug-in is developed as a proof-of-concept tool to demonstrate what we can do with the results of this study. However, given that the database with the security vulnerabilities can be updated without any change in the plug-in interface or the extension itself, new vulnerabilities can be added offline and continuously. This flexibility allows us to potentially open the database to the entire community of developers who are security experts. The inputs from the experts can be used to populate the database, which then can be viewed through the plug-in interface. This semi-private access to the database, while it remains open, can influence developers to voluntarily share their knowledge. The overall tool, i.e., the extension and the database then can be further improved based on inputs from the developers. This approach is consistent with previous study, e.g., Subramanian et al. [19] developed a browser plug-in as a proof-of-concept tool to show that code examples from Stack Overflow can be automatically included into Javadocs, but left the overall effectiveness of the tool as future work. In addition, we have also open-sourced the entire code base of the developed plug-in to promote its extensions by the community [17].

# 7 THREATS TO VALIDITY

We now discuss the threats to validity of our study following the guidelines for case study research.

Construct validity threats: Concern the relation between theory and observation. In our study, threats to the construct validity are mainly due to measurement errors.

In our study, we manually analyzed 2056 C++ code snip-pets. The code snippets are selected from 2056 code clone groups. Together, the clone groups contain all of the 72,483 code snippets in our study. Each of the 72,483 code snippets belongs to one of the 2056 clone groups. We produced the clone groups using SourcererCC. We use the state of the art clone detection tool, SourcererCC [60], to identify clones between Stack Overflow C++ code snippets. We applied tokenization at the file level in SourcererCC and set the similarity degree to 100% to find all the C++ code snippets in Stack Overflow that are exactly similar to each other (i.e., Type-1 clones). We left all the other settings of SourcererCC to their default values. Therefore, the reduction of code snippets produces no threat to our analysis. In other words, our use of SourcererCC has 100% clone detection accuracy for Type-1 with no false negatives. Therefore, when a rep-resentative of a group is reviewed for a vulnerability, the result is valid for the whole sample to which it belongs. Any finding obtained using the representatives of the groups generalizes to the whole dataset without loss of accuracy.

Since SourcererCC does not support block-level tokeniza-tion for C++ code, we designed a rule based method to track vulnerable C++ code snippets from Stack Overflow to GitHub. To evaluate this method, we randomly selected five GitHub files with migration from each of the 69 vulnerable answers; i.e., a total of 296 GitHub files and reviewed them manually. We found the approach to be 78.72% accurate. Another concern is related to false negatives that Syntaxnet [78] may have produced. Although the limitations of these different techniques may have resulted in us missing some vulnerable C++ code migrated to a GitHub project, they do not pose a threat to the validity of our findings since all vulnerability migrations reported in this paper were verified manually by multiple code reviewers. The number of vulnerability migrations reported in this paper constitutes a lower bound. There is likely much more vulnerability migrations from Stack Overflow to GitHub projects.

Internal validity threats: To avoid any misrepresentation of the information contained in the SOTorrent dataset used in this paper, we took care to remove migrated Stack Over-flow code snippets with missing GitHub links, as well as posts that were not correctly tagged. Removing these code snippets from our analysis does not pose a threat to the validity of our findings since as we mentioned above, the number of vulnerability migrations reported in this paper constitutes a lower bound. We carefully verified manually all the vulnerability migrations reported in this paper.

External validity threats: Concern the possibility to gener-alize our results. The findings reported in this paper were obtained by analyzing Stack Overflow. They therefore may not generalize to other Q&A websites. However, since the GitHub developers that we consulted did not refute the validity of any of the vulnerabilities that we reported to them and the fixes that we recommended, we believe that the results reported in this paper are strong. They offer a reliable starting point for further studies on the preva-lence and propagation of vulnerable code snippets from Q&A websites. Another potential threat to external validity concerns the generalisability of the feedback collected from developers. In fact, while recommending fixes to GitHub developers, we conducted a small survey to collect their opinion about the quality of our recommendations and the potential benefits of automatic tool supports that could warn users about the presence of vulnerabilities in code shared on Stack Overflow. Since the focus of our work was not the survey but rather the understanding of the preva-lence and propagation of vulnerable C++ code snippets from Stack Overflow to GitHub, we did not conduct a large scale survey. Nevertheless, we followed existing literature in Software Engineering [51], [52], [79] while designing our survey and the options for Likert scale [51], [52], [79], as well as during the development of our browser plug-in (that builds on the findings of the survey). Browser plug-ins are commonly developed in software engineering research projects to exemplify how results can be leveraged in practice, e.g., a browser plug-in was used to produce live software documentation from Stack Overflow [19].

## 8 Conclusion

**Summary.** In this paper, we have analyzed vulnerabilities in C++ code snippets shared on Stack Overflow and their migration to GitHub projects. This is the first study that examines the security issues of C++ code examples shared on Stack Overflow. We have investigated security vulnerabilities in the C++ code snippets shared on Stack Overflow over a period of 10 years. From the 72,483 reviewed code snippets used in at least one project hosted on GitHub, we found a total of 99 vulnerable code snippets categorized into 31 types. Bad coding practices, improper check for unusual or exceptional conditions and improper input validation were the most prevalent types of vulnerabilities. The 99 vulnerable code snippets found in Stack Overflow were reused in a total of 2859 GitHub projects. Information about the detected vulnerabilities were presented to developers of the studied GitHub projects. Although they acknowledged the vulnerabilities, many of them are still not corrected today.

**Implications.** Stack Overflow like other crowd-sourced platforms is designed to stimulate knowledge exchanges between developers. However, this platform is not equipped with a robust mechanism to ensure the good quality of answers and code snippets exchanged by its users. The incentive system (e.g., reputation, badges, upvotes, downvotes) in Stack Overflow was designed partly to encourage users to share quality contents. However, this approach works only when each user is responsible and–or knowledgeable enough about every details of the shared knowledge, which given the complex nature of software development is a difficult task. Our survey of software developers who reused vulnerable code snippets from Stack Overflow provided us with ideas for tools and techniques that can assist developers reusing code. For example, the respondents asked for both offline and online tools to inform them of any potential vulnerability in a shared code example. Consequently, we developed a browser extension that can warn developers of such vulnerable code snippets in Stack Overflow. In summary, software practitioners (e.g., developers), researchers, and Stack Overflow can benefit from our study results as follows: 1) The developers can use our developed browser extension to stay aware of potential vulnerabilities in the shared code, 2) Software engineering research can further extend our findings to analyze the diverse security aspects in the shared code and to ensure that such compromised code snippet are not included in tools built to support software development activities (e.g., tools recommending high quality posts, answers to an unanswered question, and so on) [21], [22], [23], [24]. 3) Finally, Stack Overflow can introduce a new security-focused incentive system to improve the knowledge sharing process, e.g., introduction of security badges.

**Future Work.** Our future work focuses on replicating the findings of this study to other domains (e.g., other programming languages) and venues (e.g., other crowd-sourced platforms). Based on the obtained results, we will develop new tools and techniques to promote the sharing of secured code examples, to educate developers about existing vulnerabilities, and recommend them better alternatives to an existing vulnerable code snippet.

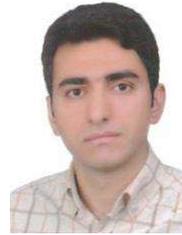

**Morteza Verdi** is an M.S. graduate of Shiraz University. He obtained his BS degree in Information Technology in 2016 from Birjand University of Technology. In December 2019, he earned his M.S. in Cyber Security from Shiraz University, where he worked under supervision of Professor Ashkan Sami. His thesis was to find vulnerability migrations between crowd-source code sharing platforms and software reposito-ries. His research interests are software engi-neering, cyber-security and artificial intelligence.

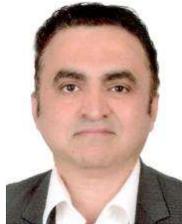

**Ashkan Sami** is an Associate Professor of Computer Science and Software Engineering at Shiraz University and National Elite's Founda-tion Professor since 2019. Ashkan teaches and conducts research on Cyber Security, Empirical Software Engineering, Applied AI and Data Science. He obtained his B.S. from Virginia Tech; U.S.A. and PhD in 2006 from Tohoku University, where his PhD became a Japanese national project and earned him a tenured faculty position at Tohoku University; Japan. He has led various interdisciplinary and transdisciplinary research teams which focuses on themes of current social problems to, create products or services and publishes in quality venues. He has published in various high quality venues like Empirical Software Engineering, MSR, IEEE Transactions on Sustainable Energy, Engineering Applications of Artificial Intelli-gence, Journal of Process Control and Gene. His current work on sys-tem and software security has been presented in media outlets like BBC Technology, The Register and professional sites like Stack Exchange blogs. In 2017, his joint applied research project won recognition as a nation's applied research project of the year. Dr. Sami has advised more than 100 M.Sc. and Ph.D. students. Web page: ashkan.synegy.ir

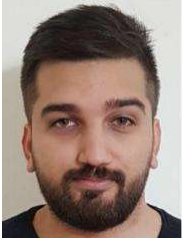

**Jafar Akhondali** is an M.S. student at Shiraz University majoring in Cyber Security. He re-ceived his B.S. from Shahid Chamran Univer-sity in 2017. His research interests are software security, wireless sensor networks and machine learning. Jafar has been working on finding se-curity problems in Stack Overflow since 2018 right after entering his M.S. program at Shiraz University. He has been actively involved in in-dustrial projects on various aspects of Software Engineering and Cyber Security for many years.

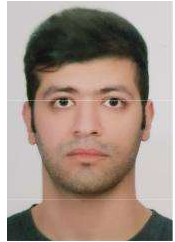

**Alireza Karami Motlagh** received his B.Sc. in Software Engineering from Shahid Chamran University of Ahvaz in 2017. He is currently an M.Sc. student at Shiraz University majoring in Artificial Intelligence. His major research inter-ests are Software Security, System Resource Management and Machine Learning. He has professional experience as a security analyst and code reviewer.

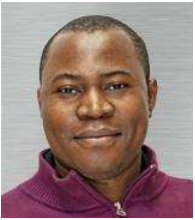

**Foutse Khomh** is a Full Professor of Soft-ware Engineering at Polytechnique Montreal´ and FRQ-IVADO Research Chair on Software Qual-ity Assurance for Machine Learning Applica-tions. He received a Ph.D in Software Engineer-ing from the University of Montreal in 2011, with the Award of Excellence. He also received a CS-Can/Info-Can Outstanding Young Computer Sci-ence Researcher Prize for 2019. His research interests include software maintenance and evo-lution, machine learning systems engineering, cloud engineering, empirical software engineering, and software ana-lytic. His work has received three ten-year Most Influential Paper (MIP) Awards, and five Best/Distinguished paper Awards. He has served on the program committees of several international conferences including FSE, ICSM(E), SANER, MSR, ICPC, SCAM, ESEM and has reviewed for top international journals such as JSS, EMSE, TSC, TSE and TOSEM. He is program chair for Satellite Events at SANER 2015, program co-chair of SCAM 2015, ICSME 2018, PROMISE 2019, and ICPC 2019, general chair of ICPC 2018, SCAM 2020, and SANER 2020. He is on the steering committee of SANER (chair), MSR, PROMISE, ICPC (chair), and ICSME(vice-chair). He initiated and co-organized the Software Engineering for Machine Learning Applications (SEMLA) symposium (https://semla.polymtl.ca/) and the RELENG (Re-lease Engineering) workshop series (http://releng.polymtl.ca). He is on the editorial board of multiple international journals, e.g., IEEE Soft-ware, Wiley's Journal of Software: Evolution and Process. Web page: http://khomh.net/.

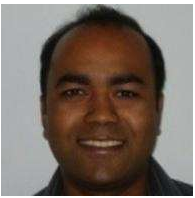

**Gias Uddin** is an Assistant Professor at the University of Calgary. He is the director of "Data Intensive Software Analytics (DISA)" lab at the university, where he leads initiatives on data in-tensive software systems. Prior to that he was a Senior Data Scientist at the Bank of Canada (the central bank) and a software engineer at IBM Watson Analytics. He completed a PhD from McGill University (2018) and a Master's from Queen's University, Canada (2008). He has worked full-time at four Industrial companies from 2008 till July 2020. In various increasingly senior roles in the Industry, he has designed and developed software using innovative machine learning techniques. The software are deployed successfully in production and are being used by thousands of millions of users. Most of his PhD works were completed while working full time at the Industry, a rewarding experience that has shaped his research on the development of practical and innovative solutions to Industrial and real-world problems. Specifically, his research focuses on the engineering of intelligent AI-driven software systems by harnessing diverse and hetero-geneous knowledge sources that can address critical problems in tech-nical, social, and organizational contexts. As such, his research work often lies at the intersection of software engineering, machine learning, natural language processing, human computer intersection, and social science. He has published papers in peer-reviewed topmost confer-ences and journals in software engineering. His recent paper at the 32nd IEEE/ACM Automated Software Engineering Conference (ASE) was nominated for a best paper award. Website: https://giasuddin.ca/